\documentclass{PoS}
\usepackage[]{units}

\title{Performance of Silicon Photomultipliers for the Dual-Mirror Medium-Sized Telescopes of the Cherenkov Telescope Array }

\ShortTitle{SiPMs for the Medium-Sized Telescopes of CTA}

\author{
\speaker{Jonathan Biteau}, David Chinn, Dennis Dang, Kevin Doyle, Caitlin A. Johnson, David A. Williams, for the CTA Consortium\footnote{Full consortium author list at http://cta-observatory.org}\\
\llap{}Santa Cruz Institute for Particle Physics and Department of Physics\\
University of California, Santa Cruz, USA\\
E-mail: \email{jbiteau@ucsc.edu}, \email{dzchinn@ucsc.edu}, \email{ddang3@ucsc.edu}, \email{kesdoyle@ucsc.edu}, \email{caajohns@ucsc.edu}, \email{daw@ucsc.edu}}

\abstract{Gamma-ray observations in the very-high-energy domain ($E > \unit[30]{GeV}$) can exploit the imaging of few-nanosecond Cherenkov flashes from atmospheric particle showers. Photomultipliers have been used as the primary photosensors to detect gamma-ray induced Cherenkov light for the past 25 years, but they are increasingly challenged by the swift progress of silicon photomultipliers (SiPMs). We are working to identify the optimal photosensors for medium-sized Schwarzschild-Couder telescopes (SCT), which are proposed to contribute a significant fraction of the sensitivity of the Cherenkov Telescope Array in its core energy range. We present the capabilities of the latest SiPMs from the Hamamatsu, SensL, and Excelitas companies that we have characterized in our laboratory, and compare them to the SiPMs equipping the prototype SCT camera that is under construction.}

\FullConference{The 34th International Cosmic Ray Conference,\\
		30 July- 6 August, 2015\\
		The Hague, The Netherlands}

\begin{document}

\section{Silicon Photomultipliers for Imaging Atmospheric Cherenkov Telescopes}

Imaging atmospheric Cherenkov telescopes, such as the current-generation H.E.S.S., MAGIC, and VERITAS, observe the sky in the very-high-energy domain (VHE, $E > \unit[30]{GeV}$) to study cosmic accelerators, constrain the nature of dark matter, and seek new physics such as violation of Lorentz invariance or coupling of gamma rays with new particles \cite{2013APh....43....3A}. The gamma-ray signal is reconstructed from atmospheric showers yielding $\sim\unit[10]{ns}$-duration Cherenkov flashes in the wavelength range $\unit[290-600]{nm}$ (full width at half maximum, FWHM), with contamination from the night sky background (NSB), particularly above $\unit[550]{nm}$. H.E.S.S., MAGIC, and VERITAS reconstruct Cherenkov signals using photomultiplier tubes (PMTs), which reach $\unit[25-35]{\%}$ photodetection efficiencies around $\unit[400]{nm}$ \cite{2013SPIE.8621E..06M}.

The Cherenkov Telescope Array (CTA) is a project under development aimed at reaching ten times improved gamma-ray sensitivity with respect to current observatories \cite{2011ExA....32..193A}. Multiple types of telescopes will coexist on site to optimize the sensitivity over more than three decades in energy while maintaining a sensible cost. We are part of an effort to develop medium-sized ($\sim\unit[10]{m}$-diameter primary mirror) Schwarzschild-Couder telescopes (SCT in the following), which feature a dual-mirror design enabling the development of a compact camera ($\sim\unit[1]{m}$-diameter). Such cameras will be equipped with silicon photomultipliers (SiPMs), whose capabilities for gamma-ray astronomy have already been demonstrated by the FACT experiment \cite{2014JInst...9P0012B}, and which are now seriously challenging PMTs in cost and performance.   

\section{Performance of recent SiPMs from Hamamatsu, SensL, and Excelitas}

The overall work on the camera built for the SCT prototype is reported elsewhere \cite{ICRC_Nepomuk}. Ultimately, the SCT cameras would be equipped with $177\times64$ pixels of approximately $\unit[6.5\times6.5]{mm^2}$ size, where the dimensions include both the active silicon area and the pitch between pixels. The prototype camera currently uses SiPMs from the Hamamatsu corporation, model S12642-0404PA-050(X), which we refer to in the following as the ``prototype devices."  These are tiles of sixteen $~\unit[3\times3]{mm^2}$ chips grouped by four to form camera image pixels, which exploit the through silicon via (TSV) technology yielding a high packing fraction in the focal plane. These devices, chosen in fall 2013 for the prototype, have now been exceeded in performance by the latest products from the Excelitas, Hamamatsu, and SensL corporations. While the choice of the final device for the construction phase will not be made before 2016, we compare in the following the latest SiPMs to the prototype device, and provide guidelines towards an optimal SCT photodetector.

\subsection{Pulse Shape}

\begin{figure}[t]
\centering
\includegraphics[width=.8\textwidth]{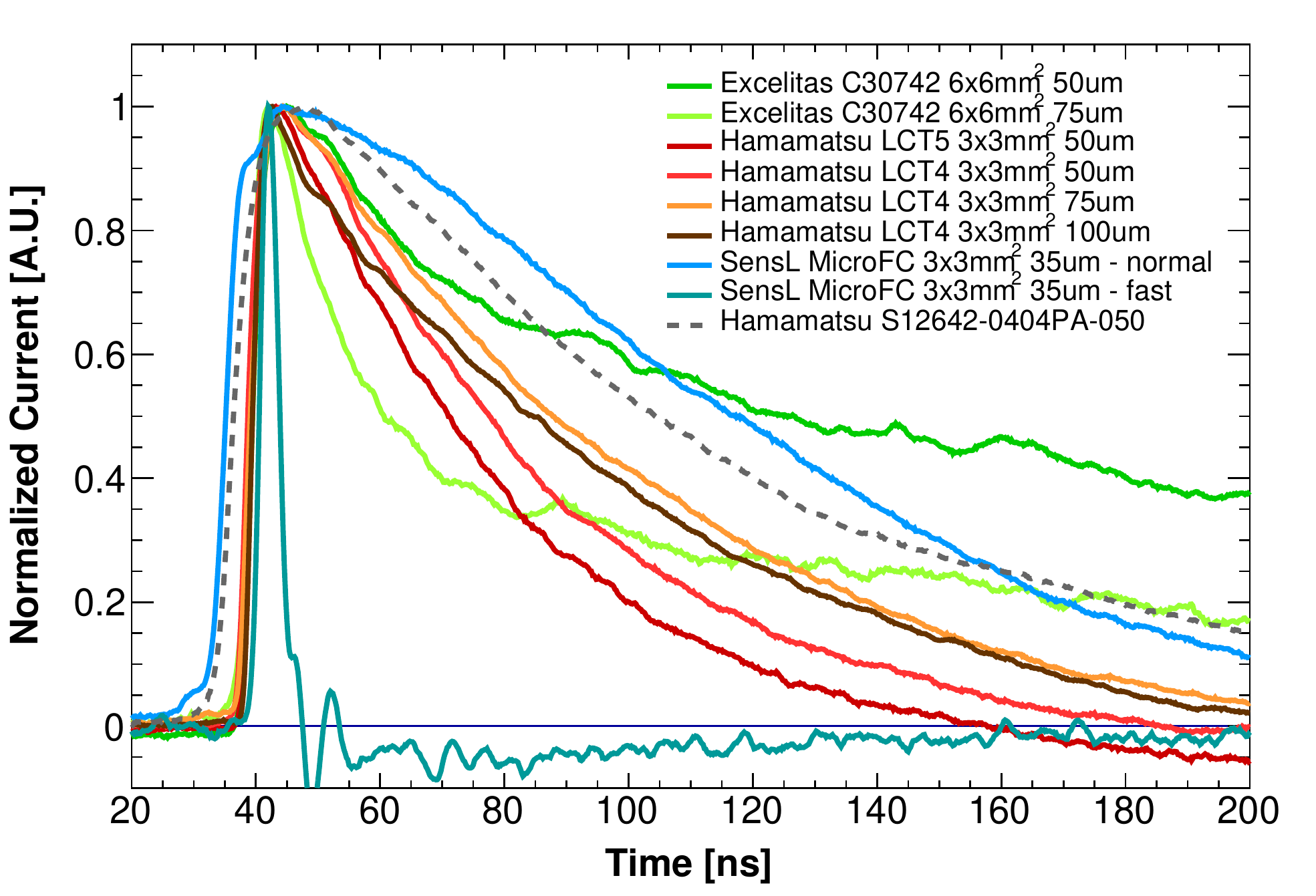}
\caption{Normalized pulse shape of the tested devices compared to that of the prototype devices (gray dashed curve). No electronic shaping is used.}
\label{fig1}
\end{figure}

The SiPM concept is fundamentally different from PMTs' design. While for the latter, pixels exploit a single photocathode, the pixels of the former are made of thousands of microcells, with typical sizes ranging from $\unit[50]{\mu m}$ up to $\unit[100]{\mu m}$. The pulses resulting from the triggering of microcells are shown for various SiPMs in Fig.~\ref{fig1}. We studied two Excelitas-C30742 devices of $\unit[6\times6]{mm^2}$ with $\unit[50]{\mu m}$ (serial number G5213) and $\unit[75]{\mu m}$ (serial number G5219) cells, respectively, one MicroFC-SMTPA SensL device ($\unit[3\times3]{mm^2}$) with the active area of microcells having $\unit[35]{\mu m}$ width, corresponding to $\sim\unit[50]{\mu m}$ when including the pitch between cells, three Hamamatsu LCT4 devices ($\unit[50]{\mu m}$, $\unit[75]{\mu m}$, and $\unit[100]{\mu m}$ cell pitch), and one Hamamatsu LCT5 device ($\unit[50]{\mu m}$ cell pitch). All the SiPMs we tested show $\unit[10-90]{\%}$ rise times between $2$ and $\unit[6]{ns}$, even when using the ``normal" output of the SensL device, which shows a rise time typically two to three times larger than the ``fast" output. The prototype devices show a rise time of $\sim\unit[7]{ns}$, electronic shaping yielding a FWHM of $\unit[10]{ns}$ \cite{Otte201585}. These values are on the order of the duration of the Cherenkov flash. We can thus conclude that SiPMs' time response is not a limiting factor for the SCT camera.

\subsection{Rate}

The breakdown voltage of a SiPM determines a switch in the behavior of its microcells, from proportional to Geiger mode. Above breakdown, the device can self trigger on thermal noise, with an increasing  dark-count rate with larger overvoltage (difference between operating and breakdown voltage). SensL, Hamamatsu, and Excelitas devices show typical breakdown voltages at $\unit[25]{^\circ C}$ of $\unit[25]{V}$, $\unit[50]{V}$, and $\unit[90]{V}$, respectively. These values slightly depend on temperature with typical variations of $\unit[0.1]{\%}$ per degree Celsius. Note that the temperature of the SCT camera is envisioned to be controlled within $\unit[0.01]{^\circ C}$ using Peltier modules.

\begin{figure}[h]
\centering
\includegraphics[width=.8\textwidth]{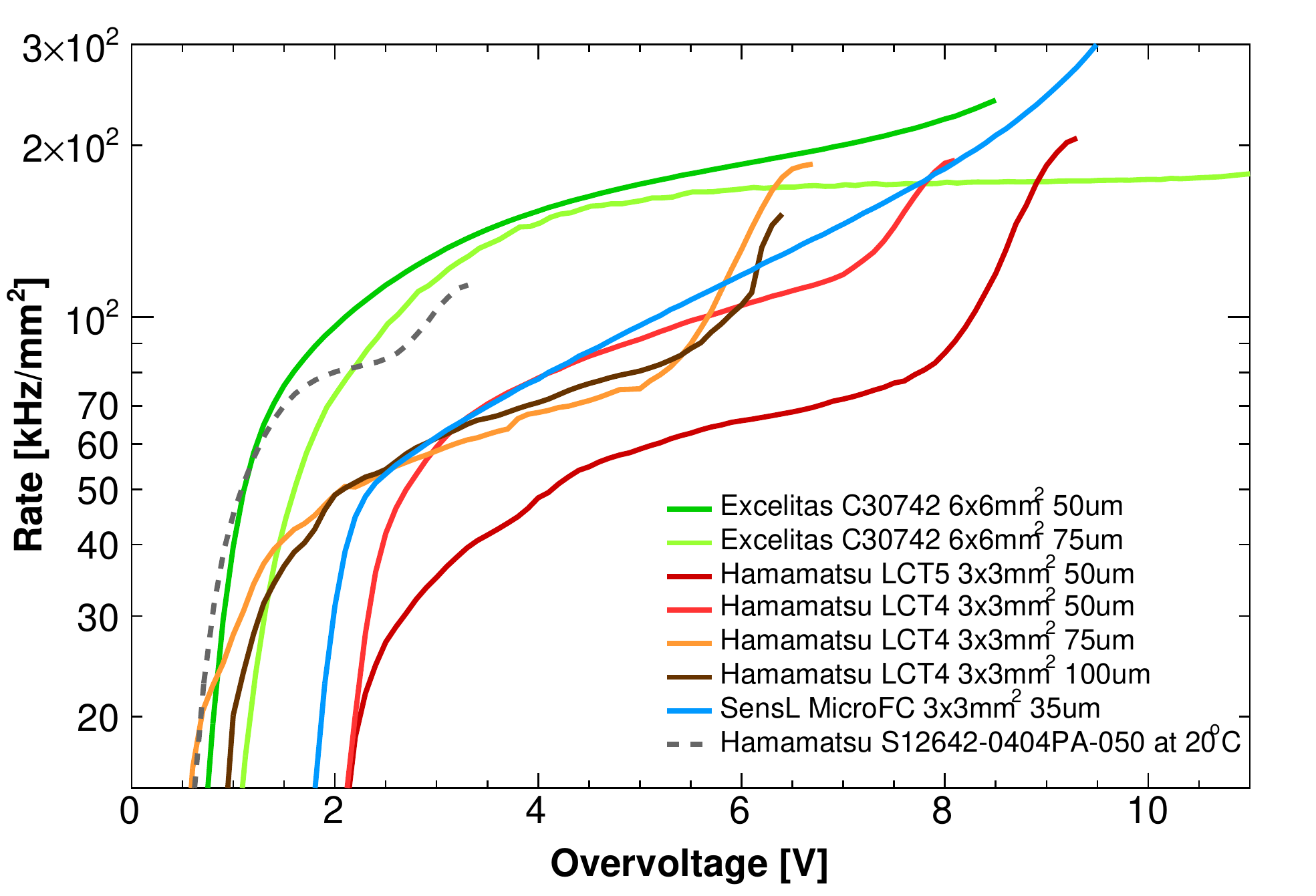}
\caption{Dark rate of the tested devices as a function of overvoltage, compared to the prototype devices (gray dashed curve). All of the measurements are at $\unit[25]{^\circ C}$, except for the prototype devices.}
\label{fig2}
\end{figure}

\begin{figure}[h!]
\centering
\includegraphics[width=.8\textwidth]{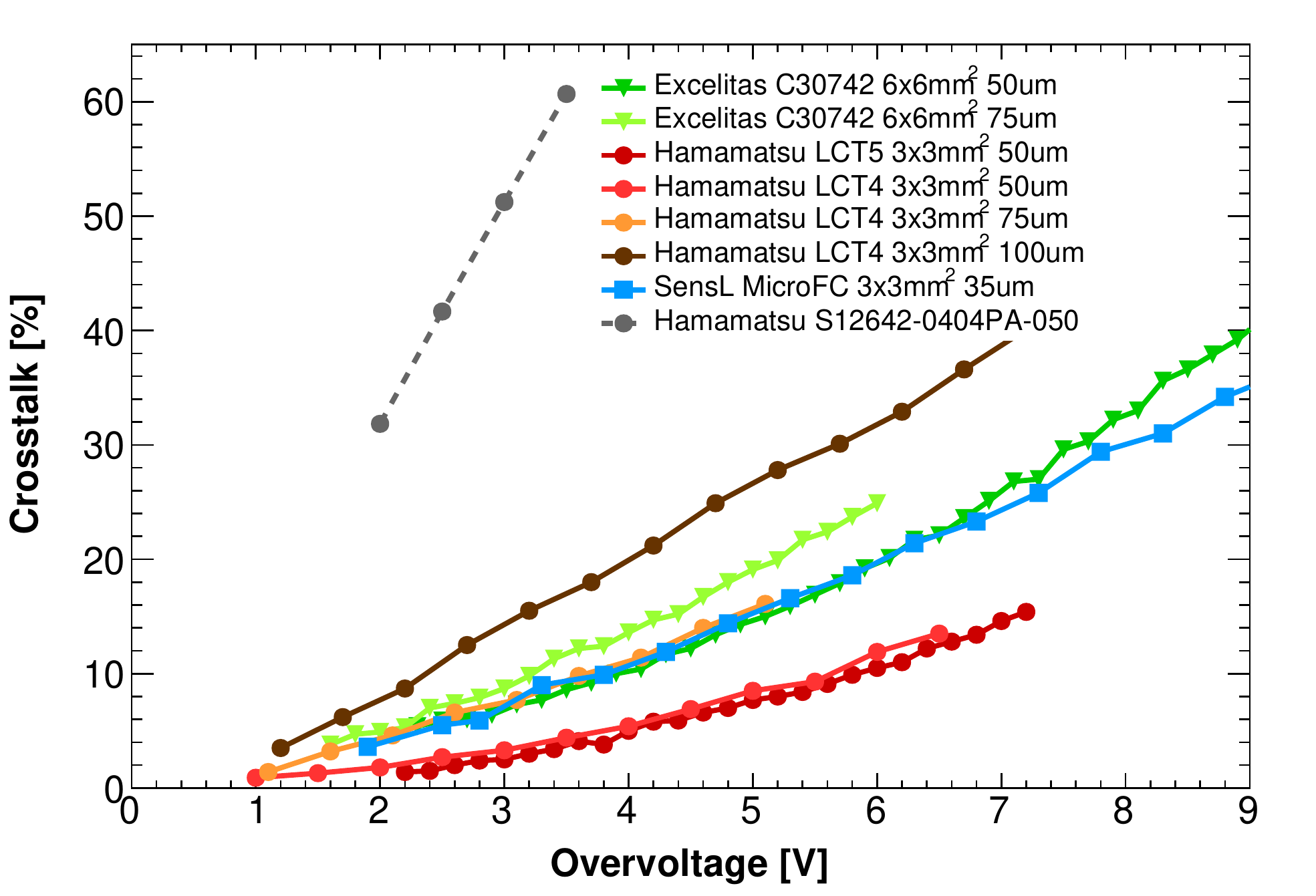}
\caption{Crosstalk of the tested devices as a function of overvoltage, compared to the prototype devices (gray dashed curve).}
\label{fig3}
\end{figure}

Figure~\ref{fig2} shows the rate of the tested devices measured at $\unit[25]{^\circ C}$ in dark conditions. At typical operating voltage ($\unit[2-8]{V}$ above breakdown, depending on the device), the rate is about $\unit[0.1]{MHz/mm^2}$ within a factor of two. This is an order of magnitude below the NSB rate, making the physical background the limiting factor for a SiPM-based camera. Cooling the SiPM device significantly reduces the thermal noise, with a dark rate divided by more than a factor of 5 when going from $\unit[25]{^\circ C}$ to $\unit[5]{^\circ C}$. We conclude, as for pulse shape, that the dark rate of SiPMs is not a limiting factor for the SCT camera.

\subsection{Optical Crosstalk}

The microcells of SiPMs are not entirely independent from each other. Dark or physical triggers generate optical photons that can in turn trigger neighboring cells. We characterize this phenomenon, know as optical crosstalk, using the charge distribution of thermal triggers measured in dark conditions. The one photoelectron events can easily be singled out from their higher-charge counterparts using the excellent single-photoelectron resolution of SiPMs. 

\begin{figure}[b]
\centering
\includegraphics[width=.8\textwidth]{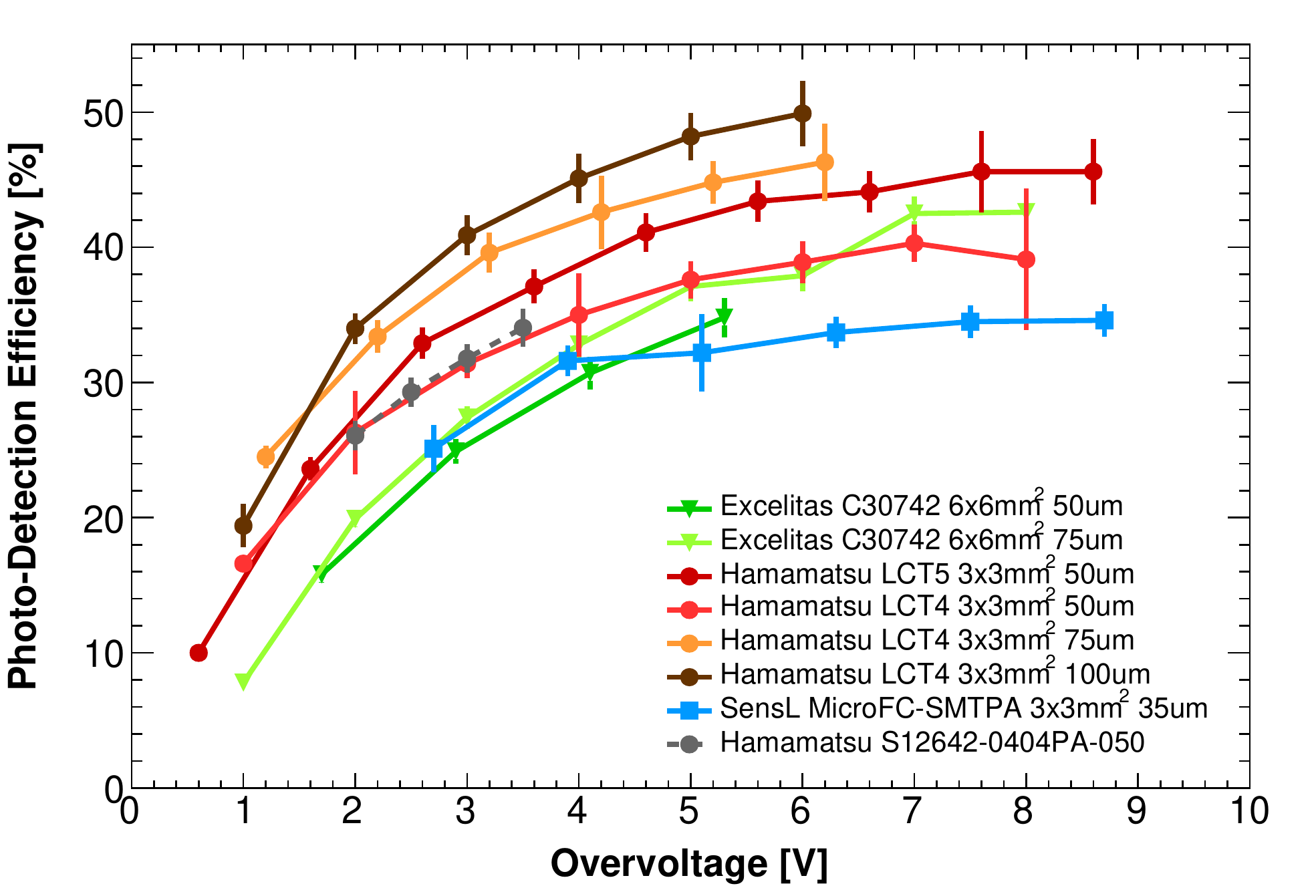}
\caption{PDE of recent devices at $\unit[400]{nm}$ as a function of overvoltage. These curves can be compared to the PDE of the prototype device shown in dashed gray.}
\label{fig4}
\end{figure}

Figure~\ref{fig3} shows the crosstalk of tested devices as a function of overvoltage. As shown by our measurements on Excelitas and Hamamatsu devices, crosstalk  steadily increases with microcell size. This can be understood as the result of the increased gain and capacitance of larger-size microcells, yielding an increased optical emission traveling to neighboring cells. We note that the $\unit[50]{\mu m}$-cell LCT4 and LCT5 devices show comparable crosstalk at similar overvoltage. 

\subsection{Photodetection Efficiency}

Photodetection efficiency (PDE) characterizes the probability that a device triggers on an incoming photon. For a SiPM, PDE includes the transmissivity of the coating, the probability to hit the active area versus dead material between microcells (filling factor), and the probability to initiate an electron-hole production, often referred to as quantum efficiency. We measure PDE with an optical setup focusing light pulses on the SiPM and on a control PIN diode, using the zero Poissonian statistics method as in \cite{2013SPIE.8852E..0KB}.

The PDEs of the Excelitas, Hamamatsu, and SensL devices at $\unit[400]{nm}$ are shown in Fig.~\ref{fig4}. LCT4 Hamamatsu devices and Excelitas devices show an increasing PDE with increasing microcell size, which is roughly compatible with the increasing filling factor. The LCT5 $\unit[50]{\mu m}$-cell device shows an enhanced PDE with respect to its LCT4 analog, with maximum values of about $\unit[45]{\%}$ and $\unit[40]{\%}$, respectively. All the tested devices show maximum PDEs between $30$ and $\unit[50]{\%}$, comparable to or better than multi-anode PMTs with similar-size pixels. 

\section{Searching for the Best SiPM for CTA Medium-Sized Telescopes}

\begin{figure}[h]
\centering
\includegraphics[width=.8\textwidth]{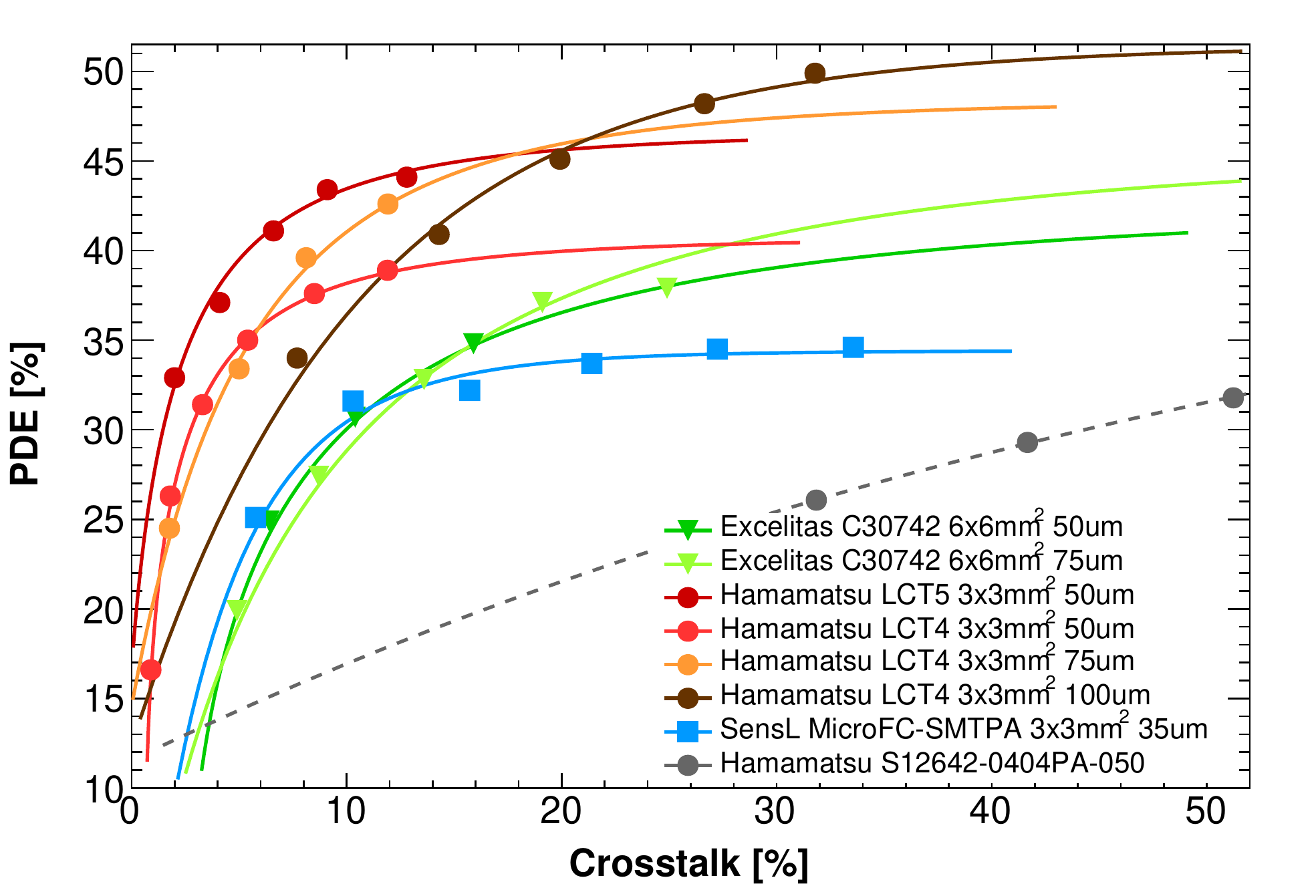}
\caption{Photodetection efficiency of recent devices at $\unit[400]{nm}$ as a function of their optical crosstalk; the dashed gray curve indicating the prototype devices. Points show measurements in overlapping overvoltage ranges of PDE and crosstalk measurements. Curves result from fits of the PDE and crosstalk data as a function of overvoltage.}
\label{fig5}
\end{figure}

We have shown that the pulse shape and dark rate of current-generation SiPMs are not limiting factors for imaging atmospheric Cherenkov telescopes. While faster rise times and lower dark rates are always desirable, PDE and crosstalk are the most important characteristics affecting the detectability of a gamma-ray shower. 
The data showing the PDE as a function of crosstalk (both of which are increasing functions of overvoltage) for particular devices are shown with colored points and curves in Fig.~\ref{fig5}. The former are shown for the range of overvoltages where we have measurements of both crosstalk and PDE. The curves result from fits of second-order polynomials to the crosstalk data and of ``saturation" functions (e.g. $1-\exp(-x)$) to the PDE data. 

\begin{figure}[h]
\centering
\includegraphics[width=.8\textwidth]{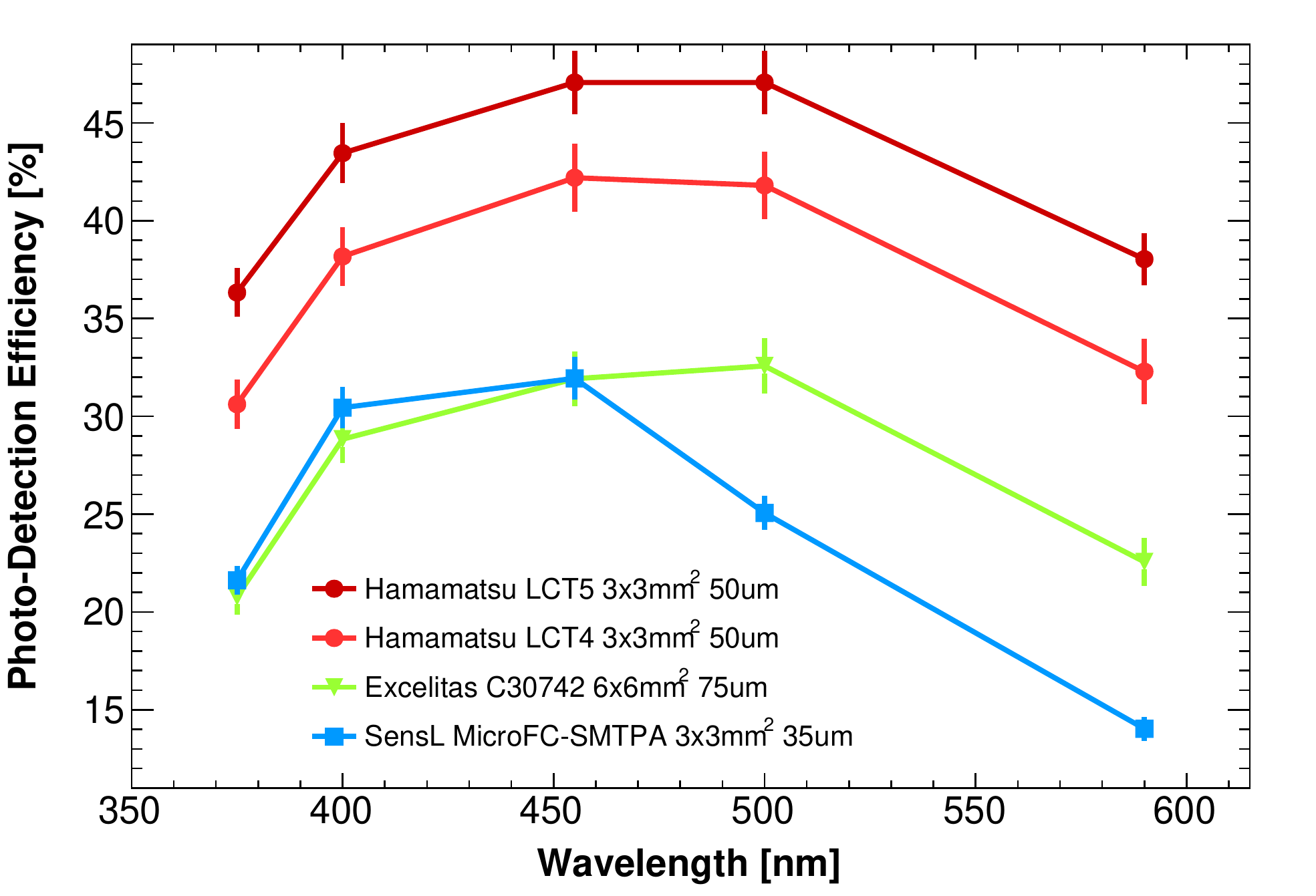}
\caption{PDE of four devices as a function of wavelength at $\unit[10]{\%}$ crosstalk.}
\label{fig6}
\end{figure}

A higher PDE results in a higher reconstruction rate of Cherenkov photons, while a higher crosstalk degrades the resolution on the charge effectively measured from a single photon. Improved SiPM capabilities for gamma-ray astronomy thus lie in the upper left corner of Fig.~\ref{fig5}. We report in Table~\ref{tab1} the PDE of the tested devices estimated at three operating points corresponding to $\unit[5]{\%}$, $\unit[10]{\%}$, and $\unit[20]{\%}$ crosstalk, respectively. The latest Hamamatsu devices show a PDE close to $\unit[40]{\%}$ at a $\unit[10]{\%}$ crosstalk, while the SensL and Excelitas devices are closer to $\unit[30]{\%}$. This corresponds to an improvement by about a factor of two with respect the prototype devices operated at a similar crosstalk level (note though that preliminary simulations favor a high-crosstalk operating point for these devices, corresponding to a PDE close to $\unit[30]{\%}$). In addition to PDE amplitude as a function of crosstalk, the comparison of devices' performance should factor in their response as a function of wavelength, such as shown in Figure~\ref{fig6}. The PDEs of Excelitas and Hamamatsu devices show maxima between 450 and $\unit[500]{nm}$, while the peak is located between $400$ and $\unit[450]{nm}$ for the SensL device. Assuming a constant mirror reflectivity as a function of wavelength, Hamamatsu and Excelitas devices would integrate a larger fraction of the red portion of the NSB, which would possibly degrade their performance with respect to SensL devices. The UV response of SiPMs between $300$ and $\unit[350]{nm}$, not probed by our current setup, is also highly relevant to gamma-ray astronomy as it corresponds to peak of the Cherenkov emission. We have recently developed a relative PDE setup exploiting the continuous emission of a deuterium lamp, which will enable probing the peak of SiPMs' wavelength response, as well as of their behavior in the near UV.\\     

\begin{table}
\centering
\begin{tabular}{l c c c}
\hline \hline
Device \qquad \qquad \qquad \quad PDE at	& $\unit[5]{\%}$	crosstalk&	$\unit[10]{\%}$ crosstalk & $\unit[20]{\%}$ crosstalk\\
\hline
Hamamatsu LCT5 $\unit[50]{\mu m}$	& $\unit[40]{\%}$ & $\unit[43]{\%}$ & $\unit[46]{\%}$ \\
Hamamatsu LCT4 $\unit[75]{\mu m}$	& $\unit[34]{\%}$ & $\unit[41]{\%}$ & $\unit[46]{\%}$ \\
Hamamatsu LCT4 $\unit[50]{\mu m}$	& $\unit[35]{\%}$ & $\unit[38]{\%}$ & $\unit[40]{\%}$ \\
Hamamatsu LCT4 $\unit[100]{\mu m}$	& $\unit[27]{\%}$ & $\unit[36]{\%}$ & $\unit[46]{\%}$ \\
SensL MicroFC-SMTPA $\unit[35]{\mu m}$	& $\unit[23]{\%}$ & $\unit[30]{\%}$ & $\unit[34]{\%}$ \\
Excelitas C30742 $\unit[50]{\mu m}$	& $\unit[20]{\%}$ & $\unit[30]{\%}$ & $\unit[37]{\%}$ \\
Excelitas C30742 $\unit[75]{\mu m}$	& $\unit[19]{\%}$ & $\unit[29]{\%}$ & $\unit[37]{\%}$\\
\hline
\end{tabular}
\caption{PDE of the tested devices at $\unit[400]{nm}$, for operating points corresponding to $\unit[5]{\%}$, $\unit[10]{\%}$, and $\unit[20]{\%}$ crosstalk. The devices are sorted by decreasing PDE at$\unit[10]{\%}$ crosstalk.}
\label{tab1}
\end{table}

Current-generation SiPMs are now highly competitive with PMTs in terms of PDE, validating the SCT concept proposed for CTA. Although lower noise is always desirable, SiPMs' dark rate is not a limiting factor, as it is about an order of magnitude below the physical noise, the NSB. SiPMs' pulses show rise times below $\unit[6]{ns}$, in compliance with the requirements of VHE astronomy. Devices from the SensL corporation have the advantage of including a fast output with a FWHM of about $\unit[4]{ns}$ for a $\unit[3\times3]{mm^2}$ device ($\unit[8]{ns}$ for $\unit[6\times6]{mm^2}$, as estimated from preliminary measurements), while electronic shaping would be needed to reduce the tails of other devices. It thus appears that PDE and crosstalk are driving the capabilities of SiPMs for CTA applications. An optimal device would work with a large PDE around $\unit[400]{nm}$ (typically above $\unit[40]{\%}$), while having a crosstalk typically smaller than $\unit[10-20]{\%}$ (upper left corner in Fig.~\ref{fig5}). Currently, only Hamamatsu LCT4 and LCT5 devices show such capabilities. For the LCT4 generation, we find that $\unit[75]{\mu m}$-cell SiPMs show better capabilities at $\unit[10]{\%}$ crosstalk than their $\unit[50]{\mu m}$ and $\unit[100]{\mu m}$ counterpart, achieving an interesting compromise between increasing PDE and crosstalk with larger microcell size. The LCT5 devices are highly promising and we look forward to testing $\unit[75]{\mu m}$-  $\unit[100]{\mu m}$-cell devices of this generation. Besides PDE and crosstalk, other criteria that will factor into the selection of a device for the SCTs will be the size of the pixels as well as the overall cost of the devices and their impact on the telescope architecture, such as the need of pulse shaping or mirror coating to reduce the red portion of the NSB.\\

We gratefully acknowledge support from the agencies and organizations listed under Funding Agencies at this website: http://www.cta-observatory.org/, and in particular the U.S. National Science Foundation award PHY-1229792 and the University of California. We also acknowledge the generosity of the SiPM manufacturers Excelitas, Hamamatsu, and SensL, who provided samples tested in this work.

\vfill

\providecommand{\href}[2]{#2}\begingroup\raggedright\endgroup

\end{document}